# AB Method of Irrigation without Water
## (Closed-loop water cycle)

**Alexander Bolonkin**
C&R, 1310 Avenue R, #F-6, Brooklyn, NY 11229, USA
T/F 718-339-4563, aBolonkin@juno.com, http://Bolonkin.narod.ru

**Abstract**

Author suggests and researches a new revolutionary idea for a closed-loop irrigation method. He offers to cover a given site by a thin closed film with controlled heat conductivity and clarity located at an altitude of 50 – 300 m. The film is supported at altitude by small additional atmospheric overpressure and connected to the ground by thin cables. Authors show that this closed dome allows full control of the weather in a given region (the day is always fine, the rain is only at night, no strong winds). The average Earth (white cloudy) reflectance equals 0.3 - 0.5. That means the Earth loses about 0.3 - 0.5 of the maximum potential incoming solar energy. The dome (having control of the clarity of film and heat conductivity) converts the cold regions to subtropics, hot deserts and desolate wildernesses to prosperous regions with a temperate climate. This is a realistic and cheap method of evaporation-economical irrigation and virtual weather control on Earth at the current time.

# Introduction
## 1. Precipitation

**1. General information about precipitation**. The extant amount of water in Earth's hydrosphere in the current era is constant. The average annual layer of Earth's precipitation is about 1000 mm or 511,000 km$^3$. 21% of this (108,000 km$^3$) falls on land and 79% (403,000 km3) on oceans. Most of it falls between latitudes 20$^o$ North and 20$^o$ South. Both polar zones collect only 4% of Earth's precipitation. The evaporation from the World-Ocean equals 1250 mm (450,000 km$^3$). 1120 mm returns back as precipitation and 130 mm by river inflow. The evaporation from land equals 410 mm (61,000 km$^3$), the precipitation is 720 mm. The land loses 310 mm as river flow to the oceans (47,000 km$^3$). These are average data. In some regions the precipitation is very different.

**2. A desert** is a landscape form or region that receives very little precipitation. Deserts are defined as areas that receive an average annual precipitation of less than 250 mm (10 in). In the Köppen climate classification system, deserts are classed as (BW).

Deserts take up one-third of the Earth's land surface. They usually have a large diurnal and seasonal temperature range, with high daytime temperatures (in summer up to 45 °C or 113 °F), and low night-time temperatures (in winter down to 0 °C; 32 °F) due to extremely low humidity. Water acts to trap infrared radiation from both the sun and the ground, and dry desert air is incapable of blocking sunlight during the day or trapping heat during the night. Thus during daylight all of the sun's heat reaches the ground. As soon as the sun sets the desert cools quickly by radiating its heat into space. Urban areas in deserts lack large (more than 25 °F/14 °C) daily temperature ranges, partially due to the urban heat island effect.

Many deserts are shielded in rain by rain shadows, mountains blocking the path of precipitation to the desert. Deserts are often composed of sand and rocky surfaces. Sand dunes called ergs and stony surfaces called hamada surfaces compose a minority of desert surfaces. Exposures of rocky terrain are typical, and reflect minimal soil development and sparseness of vegetation.

Bottomlands may be salt-covered flats. Eolian processes are major factors in shaping desert landscapes. **Cold deserts** (also known as polar deserts) have similar features but the main form of precipitation is snow rather than rain. Antarctica is the world's largest cold desert (composed of about 98 percent thick continental ice sheet and 2 percent barren rock). The largest hot desert is the Sahara. Deserts sometimes contain valuable mineral deposits that were formed in the arid environment or that were exposed by erosion.

Rain *does* fall occasionally in deserts, and desert storms are often violent. A record 44 millimeters (1.7 in) of rain once fell within 3 hours in the Sahara. Large Saharan storms may deliver up to 1 millimeter per minute. Normally dry stream channels, called arroyos or wadis, can quickly fill after heavy rains, and flash floods make these channels dangerous.

Though little rain falls in deserts, deserts receive runoff from ephemeral, or short-lived, streams fed considerable quantities of sediment for a day or two. Although most deserts are in basins with closed or interior drainage, a few deserts are crossed by 'exotic' rivers that derive their water from outside the desert. Such rivers infiltrate soils and evaporate large amounts of water on their journeys through the deserts, but their volumes are such that they maintain their continuity. The Nile River, the Colorado River, and the Yellow River are exotic rivers that flow through deserts to deliver their sediments to the sea. Deserts may also have underground springs, rivers, or reservoirs that lay close to the surface, or deep underground. Plants that have not completely adapted to sporadic rainfalls in a desert environment may tap into underground water sources that do not exceed the reach of their root systems.

Lakes form where rainfall or meltwater in interior drainage basins is sufficient. Desert lakes are generally shallow, temporary, and salty. Because these lakes are shallow and have a low bottom gradient, wind stress may cause the lake waters to move over many square kilometers. When small lakes dry up, they leave a salt crust or hardpan. The flat area of clay, silt, or sand encrusted with salt that forms is known as a playa. There are more than a hundred playas in North American deserts. Most are relics of large lakes that existed during the last ice age about 12,000 years ago. Lake Bonneville was a 52,000 kilometers² (20,000 mi²) lake almost 300 meters (1000 ft) deep in Utah, Nevada, and Idaho during the Ice Age. Today the remnants of Lake Bonneville include Utah's Great Salt Lake, Utah Lake, and Sevier Lake. Because playas are arid landforms from a wetter past, they contain useful clues to climatic change.

When the occasional precipitation does occur, it erodes the desert rocks quickly and powerfully. Winds are the other factor that erodes deserts—they are slow yet constant.

A desert is a hostile, potentially deadly environment for unprepared humans. The high heat causes rapid loss of water due to sweating, which can result in dehydration and death within days. In addition, unprotected humans are also at risk from heatstroke and venomous animals. Despite this, some cultures have made deserts their home for thousands of years, including the Bedouin, Touareg and Puebloan people. Modern technology, including advanced irrigation systems, desalinization and air conditioning have made deserts much more hospitable. In the United States and Israel, desert farming has found extensive use.

The Great Sandy Desert has nearly all its rain during from monsoonal thunderstorms or the occasional tropical cyclone rain depression. Thunderstorm days average 20-30 annually through most of the area (Burbidge 1983) although the desert has fairly high precipitation rates due to the high rates of evaporation this area remains an arid environment with vast areas of sands.

Other areas of the world, which see these rare precipitation events in drylands, are Northwest Mexico, South West America, and South West Asia. In North America in the Sonoran and Chihuahuan desert have received some tropical rainfall in the last 10 years. Tropical activity is rare in all deserts but what rain does arrive here is important to the delicate ecosystem existing.

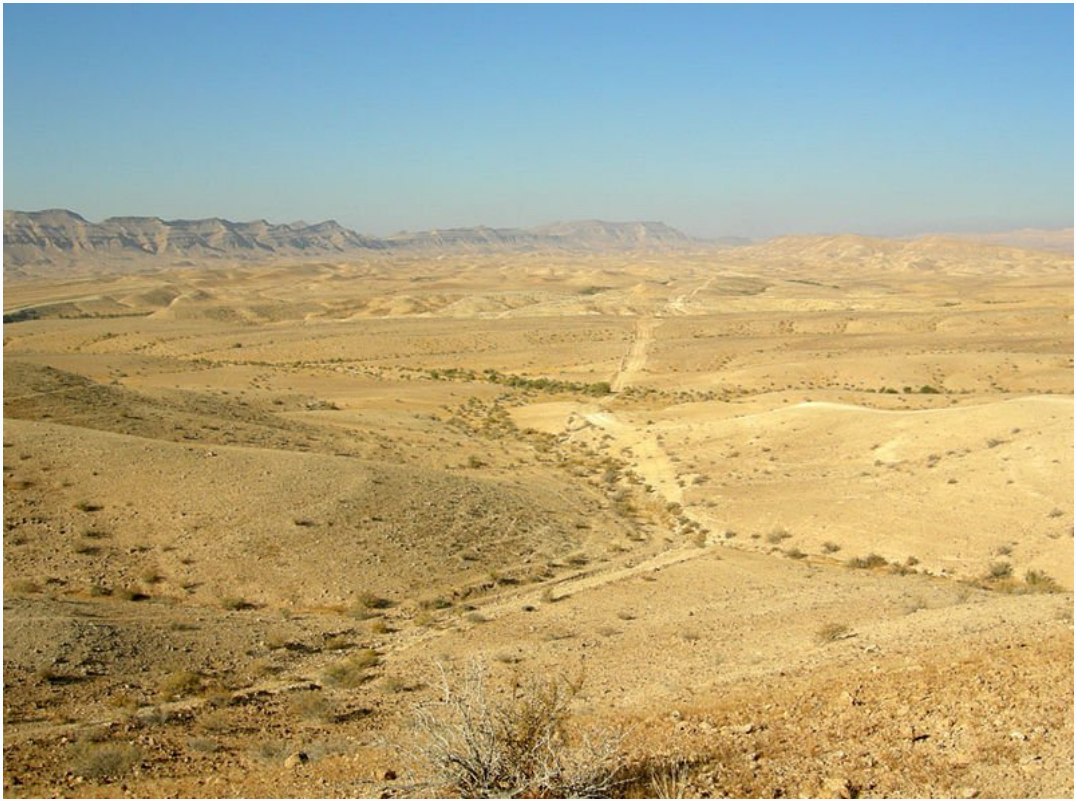

**Fig.1.** Mahktesh Gadol, an erosional basin in the Negev Desert of southern Israel.213 kb.

**3.Arid.** In general terms, the climate of a locale or region is said to be **arid** when it is characterized by a severe lack of available water, to the extent of hindering or even preventing the growth and development of plant and animal life. As a result, environments subject to arid climates tend to lack vegetation and are called xeric or desertic.

The expression 'available water' refers to water in the soil in excess to the wilting point. The air over a hot desert may actually contain substantial amounts of water vapor but that water may not be generally accessible to plants, except for very specialized organisms (such as some species of lichen). 'Lack of water' refers to use by plants. The water that is actually present in the environment may be sufficient for some species or usages (such as climax vegetation), and grossly insufficient for others. **Aridity**, the characteristic nature of arid climates, may thus depend on the use of the land. Regards to the presence of life, what is more important than the degree of rainfall is the fraction of precipitation that is not quickly lost through evaporation or runoff. Attempts to quantitatively describe the degree of aridity of a place has often led to the development of aridity indexes. There is no universal agreement on the precise boundaries between classes such as 'hyper-arid', 'arid', 'semi-arid', etc.

If different classification schemes and maps differ in their details, there is a general agreement about the fact that large areas of the Earth are considered arid. These include the hot deserts located broadly in sub-tropical regions, where the accumulation of water is largely prevented by either low precipitations, or high evaporation, or both, as well as cold deserts near the poles, where water may be permanently locked in solid forms (snow and ice). Other arid regions include areas located in the rain shadows of major mountain ranges or along coastal regions affected by significant upwelling (such as the Atacama Desert).

The distribution of aridity observed at any one point in time is largely the result of the general circulation of the atmosphere. The latter does change significantly over time through climate change. In addition, changes in land use can result in greater demands on soil water and induce a higher degree of aridity.

**4. A drought** is an extended period of months or years when a region notes a deficiency in its water supply. Generally, this occurs when a region receives consistently below average precipitation. It can have a substantial impact on the ecosystem and agriculture of the affected region. Although droughts can persist for several years, even a short, intense drought can cause significant damage and harm the local economy.

Drought is a normal, recurring feature of the climate in most parts of the world. Having adequate drought mitigation strategies in place can greatly reduce the impact. Recurring or long-term drought can bring about desertification. Recurring droughts in the Horn of Africa have created grave ecological catastrophes, prompting massive food shortages, still recurring. To the north-west of the Horn, the Darfur conflict in neighboring Sudan, also affecting Chad, was fueled by decades of drought; combination of drought, desertification and overpopulation are among the causes of the Darfur conflict, because the Arab Baggara nomads searching for water have to take their livestock further south, to land mainly occupied by non-Arab farming peoples.

According to a UN climate report, the Himalayan glaciers that are the sources of Asia's biggest rivers - Ganges, Indus, Brahmaputra, Yangtze, Mekong, Salween and Yellow - could disappear by 2035 as temperatures rise. Approximately 2.4 billion people live in the drainage basin of the Himalayan rivers. India, China, Pakistan, Bangladesh, Nepal and Myanmar could experience floods followed by droughts in coming decades. Drought in India affecting the Ganges is of particular concern, as it provides drinking water and agricultural irrigation for more than 500 million people. Paradoxically, some proposed short-term solutions to global warming also carry with them increased chances of drought.

In 2005, parts of the Amazon basin experienced the worst drought in 100 years. A 23 July 2006 article reported Woods Hole Research Center results showing that the forest in its present form could survive only three years of drought. Scientists at the Brazilian National Institute of Amazonian Research argue in the article that this drought response, coupled with the effects of deforestation on regional climate, are pushing the rainforest towards a "tipping point" where it would irreversibly start to die. It concludes that the rainforest is on the brink of being turned into savanna or desert, with catastrophic consequences for the world's climate. According to the WWF, the combination of climate change and deforestation increases the drying effect of dead trees that fuels forests fires.

**5. Tundra**. In physical geography, **tundra** is an area where the tree growth is hindered by low temperatures and short growing seasons. The term "tundra" comes from Kildin Sami *tūndâr* 'uplands, tundra, treeless mountain tract'. There are two types of tundra: Arctic tundra (which also occurs in Antarctica), and alpine tundra. In tundra, the vegetation is composed of dwarf shrubs, sedges and grasses, mosses, and lichens. Scattered trees grow in some tundra. The ecotone (or ecological boundary region) between the tundra and the forest is known as the tree line or timberline.

Arctic tundra occurs in the far Northern Hemisphere, north of the taiga belt. The word "tundra" usually refers only to the areas where the subsoil is permafrost, or permanently frozen soil. (It may also refer to the treeless plain in general, so that northern Sápmi would be included.) Permafrost tundra includes vast areas of northern Russia and Canada. The polar tundra is home to several peoples who are mostly nomadic reindeer herders, such as the Nganasan and Nenets in the permafrost area (and the Sami in Sápmi).

The Arctic tundra is a vast area of stark landscape, which is frozen for much of the year. The soil there is frozen from 25-90 cm (9.8-35.4 inches) down, and it is impossible for trees to grow. Instead, bare and sometimes rocky land can only support low growing plants such as moss, heath, and lichen. There are two main seasons, winter and summer, in the polar Tundra areas. During the winter it is very cold and dark, with the average temperature around -28 °C (-18.4°F), sometimes dipping as low as -50 °C (-58°F). However, extreme cold temperatures on the tundra do not drop as low as those experienced in taiga areas further south (for example, Russia's and Canada's lowest temperatures were recorded in locations south of the treeline). During the summer, temperatures rise somewhat, and the top layer of the permafrost melts, leaving the ground very soggy. The tundra is covered in marshes, lakes, bogs and streams during the warm months. Generally daytime temperatures during the summer rise to about 12°C (53.6°F) but can often drop to 3°C (37.4°F) or even below freezing. Arctic tundras are sometimes the subject of habitat conservation programs. In Canada and Russia, many of these areas are protected through a national Biodiversity Action Plan.

The tundra is a very windy area, with winds often blowing upwards at 48–97 km/h (30-60 miles an hour). However, in terms of precipitation, it is desert-like, with only about 15–25 cm (6–10 inches) falling per year (the summer is typically the season of maximum precipitation). During the summer, the permafrost thaws just enough to let plants grow and reproduce, but because the ground below this is frozen, the water cannot sink any lower, and so the water forms the lakes and marshes found during the summer months. Although precipitation is light, evaporation is also relatively minimal.

The biodiversity of the tundras is low: 1,700 species of vascular plants and only 48 land mammals can be found, although thousands of insects and birds migrate there each year for the marshes. There are also a few fish species such as the flat fish. There are few species with large populations. Notable animals in the Arctic tundra include caribou (reindeer), musk ox, arctic hare, arctic fox, snowy owl, lemmings, and polar bears (only the extreme north).

Due to the harsh climate of the Arctic tundra, regions of this kind have seen little human activity, even though they are sometimes rich in natural resources such as oil and uranium. In recent times this has begun to change in Alaska, Russia, and some other parts of the world.

A severe threat to the tundras, specifically to the permafrost, is global warming. Permafrost is essentially a frozen bog - in the summer, only its surface layer melts. The melting of the permafrost in a given area on human time scales (decades or centuries) could radically change which species can survive there.

Another concern is that about one third of the world's soil-bound carbon is in taiga and tundra areas. When the permafrost melts, it releases carbon in the form of carbon dioxide, a greenhouse gas. The effect has been observed in Alaska. In the 1970s the tundra was a carbon sink, but today, it is a carbon source.

**6. Permafrost**. 65% of Russian territory is permafrost. In geology, **permafrost** or **permafrost soil** is soil at or below the freezing point of water (0 °C or 32 °F) for two or more years. Ice is not always present, as may be in the case of nonporous bedrock, but it frequently occurs and it may be in amounts exceeding the potential hydraulic saturation of the ground material. Most permafrost is located in high latitudes (e.g. North and South poles), but **alpine permafrost** exists at high altitudes.

The extent of permafrost can vary as the climate changes. Today, approximately 20% of the Earth's land mass is covered by permafrost (including discontinuous permafrost) or glacial ice. Overlying permafrost is a thin *active layer* that seasonally thaws during the summer. Plant life can be supported only within the active layer since growth can occur only in soil that is fully thawed for some part of the year. Thickness of the active layer varies by year and location, but is typically 0.6–

4 m (2 to 12 feet) thick. In areas of continuous permafrost and harsh winters the depth of the permafrost can be as much as 1493 m (4510 ft) in the northern Lena and Yana River basins in Siberia.

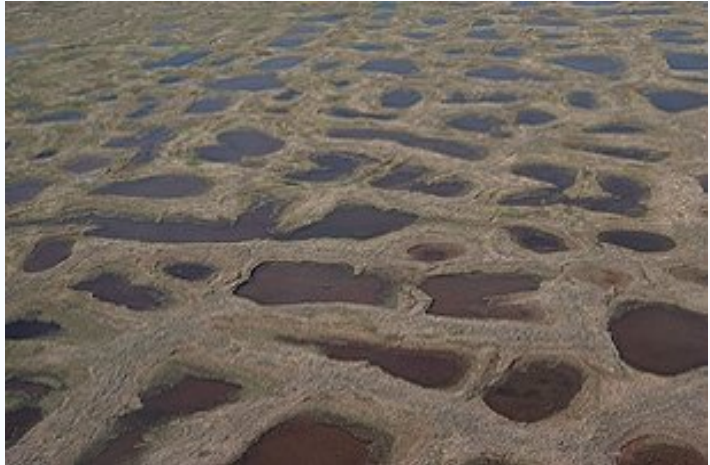

**Fig.2**. Permafrost. 36 kb

## 2. Irrigation

We see, then that more then *half* of Earth's land surface in dry or cold regions, which are by definition of these conditions unsuitable for agriculture. Many dry regions and deserts are not used because no water is available for irrigation. In the rare cases where it is practical, humanity spends a lot of money and energy to obtain fresh water, building the irrigation canals, pumping stations and distribution systems—and in the long term, still more fighting silting which literally can require rebuilding such works generation by generation.

**1. Irrigation** is the artificial application of water to the soil usually for assisting in growing crops. In crop production it is mainly used to replace missing rainfall in periods of drought, but also to protect plants against frost. Additionally irrigation helps to suppress weed growing in rice fields.[2] In contrast, agriculture that relies only on direct rainfall is sometimes referred to as dryland farming or as rain fed farming. It is often studied together with drainage, which is the natural or artificial removal of surface and sub-surface water from a given area.

Various types of irrigation techniques differ in how the water obtained from the source is distributed within the field. In general, the goal is to supply the entire field uniformly with water, so that each plant has the amount of water it needs, neither too much nor too little.

By the middle of the 20th century, the advent of diesel and electric motors led for the first time to systems that could pump groundwater out of major aquifers faster than it was recharged. This can lead to permanent loss of aquifer capacity, decreased water quality, ground subsidence, and other problems. The future of food production in such areas as the North China Plain, the Punjab, and the Great Plains of the US is threatened.

At the global scale 2,788,000 sq km (689 million acres) of agricultural land was equipped with irrigation infrastructure around the year 2000. About 68 % of the area equipped for irrigation is located in Asia, 17 % in America, 9 % in Europe, 5 % in Africa and 1 % in Oceania. The largest contiguous areas of high irrigation density are found in North India and Pakistan along the rivers Ganges and Indus, in the Hai He, Huang He and Yangtze basins in China, along the Nile river in Egypt and Sudan, in the Mississippi-Missouri river basin and in parts of California. Smaller irrigation areas are spread across almost all populated parts of the world.[16]

Irrigation gives high stability harvests which are ~ 3 – 5 times more than conventional agriculture would tend to provide.

For example, the irrigation in the Central Asia (former USSR) gives wheat equal to 65 – 70 *centners* (1 centner = 100 kg, a metric hundredweight) from 1 hectare ($10^4$ $m^2$), rice 80 – 100 centners from 1 hectare, raw-cotton 40 – 50 centners from 1 hectare. Irrigation is performed only on 16% of all active agricultural area but it produces as much of the agriculture production as all non-irrigated cultivation.

Sources of irrigation water can be groundwater extracted from springs or by using wells, surface water withdrawn from rivers, lakes or reservoirs or non-conventional sources like treated wastewater, desalinated water or drainage water. A special form of irrigation using surface water is spate irrigation, also called floodwater harvesting. In case of a flood (spate) water is diverted to normally dry river beds (wadi's) using a network of dams, gates and channels and spread over large areas. The moisture stored in the soil will be used thereafter to grow crops. Spate irrigation areas are in particular located in semi-arid or arid, mountainous regions. While floodwater harvesting belongs to the accepted irrigation methods, rainwater harvesting is usually not considered as a form of irrigation. Rainwater harvesting is the collection of runoff water from roofs or unused land and the concentration of this water on cultivated land. Therefore this method is considered as a water concentration method.

Problems in conventional irrigation:

- Competition for surface water rights.
- Depletion of underground aquifers.
- Ground subsidence (e.g. New Orleans, Louisiana)
- Buildup of toxic salts on soil surface in areas of high evaporation. This requires either leaching to remove these salts and a method of drainage to carry the salts away or use of mulch to minimize evaporation.
- Overirrigation because of poor distribution uniformity or management wastes water, chemicals, and may lead to water pollution.

The one man spent from 10 - 50 liters per day (last number for industrial counries and includes the watering of house garden). The huge amount of water request the plants. For example, the one hectare of wheat requests 2000 kL (kL is kiloliter = 1 ton), the cabbage - 8000 kL, the forest 12,000 - 15,000 kL per summer.

2. **Hydroponics** is a method of growing plants using mineral nutrient solutions instead of soil. Terrestrial plants may be grown with their roots in the mineral nutrient solution only or in an inert medium, such as perlite, gravel or Rockwool. A variety of techniques exist.

Plant physiology researchers discovered in the 19th century that plants absorb essential mineral nutrients as inorganic ions in water. In natural conditions, soil acts as a mineral nutrient reservoir but the soil itself is not essential to plant growth. When the mineral nutrients in the soil dissolve in water, plant roots are able to absorb them. When the required mineral nutrients are introduced into a plant's water supply artificially, soil is no longer required for the plant to thrive. Almost any terrestrial plant will grow with hydroponics, but some will do better than others. It is also very easy to do; the activity is often undertaken by very young children with such plants as watercress. Hydroponics is also a standard technique in biology research and teaching and a popular hobby.

Due to its arid climate, Israel has developed advanced hydroponic technology. They have marketed their system to Nicaragua, which uses it to produce more than one million pounds of peppers annually for sale abroad, including the United States.

The largest commercial hydroponics facility in the world is Eurofresh Farms in Willcox, Arizona, which sold 125 million pounds of tomatoes in 2005. Eurofresh has 256 acres under glass and represents about a third of the commercial hydroponic greenhouse area in the U.S. Eurofresh does not consider their tomatoes organic, but they are pesticide-free. They are grown in rockwool with top irrigation.

Some commercial installations use no pesticides or herbicides, preferring integrated pest management techniques. There is often a price premium willingly paid by consumers for produce which is labeled "organic". Some states in the USA require soil as an essential to obtain organic certification. There are also overlapping and somewhat contradictory rules established by the US Federal Government, so some food grown with hydroponics can be certified organic. In fact, they are the cleanest plants possible because there is no environment variable and the dirt in the food supply is extremely limited. Hydroponics also saves an incredible amount of water; it uses as little as 1/20 the amount as a regular farm to produce the same amount of food. The water table can be impacted by the water use and run-off of chemicals from farms, but hydroponics may minimize impact as well as having the advantage that water use and water returns are easier to measure. This can save the farmer money by allowing reduced water use and the ability to measure consequences to the land around a farm.

The environment in a hydroponics greenhouse is tightly controlled for maximum efficiency and this new mindset is called Soil-less/Controlled Environment Agriculture (S/CEA). With this growers can make ultra-premium foods anywhere in the world, regardless of temperature and growing seasons. Growers monitor the temperature, humidity, and pH level constantly.

Hydroponics have been used to enhance vegetables to provide more nutritional value. A hydroponic farmer in Virginia has developed a calcium and potassium enriched head of lettuce, scheduled to be widely available in April of 2007. Grocers in test markets have said that the lettuce sells "very well," and the farmers claim that their hydroponic lettuce uses 90% less water than traditional soil farming.

Advantages, disadvantages, and misconceptions:

- While removing soil-grown crops from the ground effectively kills them, hydroponically grown crops such as lettuce can be packaged and sold while still alive, greatly increasing the length of freshness once purchased.
- Solution culture hydroponics does not require disposal of a solid medium or sterilization and reuse of a solid medium.
- Solution culture hydroponics allows greater control over the root zone environment than soil culture.
- Over- and under-watering is prevented
- Hydroponics is often the best crop production method in remote areas that lack suitable soil, such as Antarctica, space stations, space colonies, or atolls such as Wake Island.
- In solution culture hydroponics, plant roots can be seen.
- Soil borne diseases are virtually eliminated.
- Weeds are virtually eliminated.
- Fewer pesticides may be required because of the above two reasons.
- Edible crops are not contaminated with soil.
- Water use can be substantially less than with outdoor irrigation of soil-grown crops.
- Hydroponics cost 20% less than other ways for growing strawberries.
- Many hydroponic systems give the plants more nutrition while at the same time using less energy and space.
- Hydroponics allow for easier fertilization as it is possible to use an automatic timer to fertilize the plants.

- It provides the plant with balanced nutrition because the essential nutrients are dissolved into the water-soluble nutrient solution.

- If timers or electric pumps fail or the system clogs or springs a leak, plants can die very quickly in many kinds of hydroponic systems.
- Hydroponics usually requires a greater technical knowledge than **geoponics**.
- For the previous two reasons and the fact that most hydroponic crops are grown in greenhouses or controlled environment agriculture, hydroponic crops are usually more expensive than soil-grown crops.
- Solution culture hydroponics requires that the plants be supported because the roots have no anchorage without a solid medium.
- The plants will die if not frequently monitored while soil plants do not require such close attention.

Hydroponics has been widely misconceived as miraculous. There are many widely held misconceptions regarding hydroponics, as noted by the following facts:

- Hydroponics will not always produce greater crop yields than with good quality soil.
- Hydroponic plants cannot always be spaced closer together than soil-grown crops (**geoponics**) under the same environmental conditions.
- Hydroponic produce will not necessarily be more nutritious or better tasting than geoponics.[12]

Hydroponics will grow 30% faster and cost less. They are also proven to be healthier and more productive.

With pest problems reduced, and nutrients constantly fed to the roots, productivity in hydroponics is high, plant growth being limited by the low levels of carbon dioxide in the atmosphere, or limited light. To increase yield further, some sealed greenhouses inject carbon dioxide into their environment to help growth ($CO_2$ enrichment), or add lights to lengthen the day, control vegetative growth etc.

This technology allows for growing where no one has grown before, be it underground, or above, in space or under the oceans this technology will allow humanity to live where humanity chooses. If used for our own survival or our colonisation, hydroponics is and will be a major part of our collective future.

3. **Control of local weather**. Governments spend billions of dollars merely studying the weather. The many big government research scientific organizations and perhaps a hundred thousand of scientists have been studying Earth's weather for more than a hundred years. There are gigantic numbers of scientific works about weather control. Most of them are impractical. We cannot exactly predict weather at long period, to avert a rain, strong wind, storm, tornado, or hurricane. We cannot control the clouds, temperature and humidity of the atmosphere, nor the power of rain. We cannot make more tolerable a winter or summer. We cannot convert a cold region to subtropics, a desolate wilderness to a prosperous region. We can only observe the storms and hurricanes and approximately predict their direction of movement. It is as if all the police department did was announce which neighborhoods were infested with killers and best avoided! Every year terrible storms, hurricanes, strong winds and rains and inundations destroy thousands of houses, kill thousands of men.

## 2. DESCRIPTION AND INNOVATIONS

Our idea is a closed dome covering a local region by a thin film with controlled heat conductivity and (optionally) controlled clarity (reflectivity, albedo, carrying capacity of solar spectrum). The film is located at an altitude of ~(50 – 300 m). The film is support at this altitude by a small additional air pressure produced by ground ventilators. The film is connected to Earth's ground by cables. The cover may require double-layer film. We can control the heat conductivity of the dome cover by pumping an air between two layers of the dome cover and change the solar heating (solar radiation) by control of cover clarity. That allows selecting for different conditions (solar heating) in the covered area and by pumping air into dome. Envisioned is a cheap film having liquid crystal and conducting layers. The clarity is controlled by electric voltage. These layers, by selective control, can pass or blockade the solar light (or parts of solar spectrum) and pass or blockade the Earth's radiation. The incoming and outgoing radiations have different wavelengths. That makes control of them separately feasible and therefore possible to manage the heating or cooling of the Earth's surface under this film. In conventional conditions about 50% of the solar energy reaches the Earth surface. Much is reflected back to outer space by the white clouds. In our closed water system the rain (or at least condensation) will occur at night when the temperature is low. In open atmosphere, the Sun heats the ground; the ground must heat the whole troposphere (4 – 5 km) before stable temperature rises are achieved. In our case the ground heats ONLY the air into the dome (as in a hotbed). We have a literal greenhouse effect. That means that many cold regions (Alaska, Siberia, North Canada) may absorb more solar energy and became a temperate climate or sub-tropic climate (under the dome, as far as plants are concerned). That also means the Sahara and other deserts can be a prosperous area with a fine growing and living climate and with a closed-loop water cycle.

The building of a film dome is very easy. We spread out the film over Earth's surface, turn on the pumping propellers and the film is raised by air overpressure to the needed altitude limited by the support cables. Damage to the film is not a major trouble because the additional air pressure is very small (0.0001- 0.01 atm) and air leakage is compensated for by propeller pumps. Unlike in a space colony or planetary colony, the outside air is friendly and at worst we lose some heat (or cold) and water vapor.

**The first main innovation** of the offered dome (and main difference from a conventional hotbed, or greenhouse) is the inflatable HIGH span of the closed cover (up to 50 – 300 m). The high height of the enclosed volume aids organizing of a CLOSED LOOP water cycle - accepting of water vaporized by plants and returning this water in the nighttime when the air temperature decreases. That allows us to perform irrigation in the gigantic portion of Earth's land area that does not have enough water for agriculture. We can convert the desert and desolate wildernesses into gardens without expensive delivery of remote freshwater. The initial amount of water for water cycle may be collected from atmospheric precipitation in some period or delivered. Good soil is not a necessity-- hydroponics allows us to achieve record harvests on any soil.

**The second important innovation** is using a cheap controlled heat conductivity, double-layer cover (controlled clarity is optionally needed for some regions). This innovation allows to conserve solar heat (in cold regions), to control temperature (in hot climates). That allows to get two to three rich crops annually in middle latitudes and to convert the cold zones (Siberia, North Canada, Alaska, etc.) to good single-crop regions.

**The third innovation** is control of the cover height, which allows adapting to local climatic seasons.

**The fourth innovation** is the using cheap thin film as the high altitude cover. This innovation decreases the construction cost by thousands of times in comparison with the conventional very expensive glass-concrete domes offered by some authors for city use.

Lest it be objected that such domes would take impractical amounts of plastic, consider that the world's plastic production is today on the order of 100 million tons. If, with economic growth, this amount doubles over the next generation and the increase is used for doming over territory, at 500 tons a square kilometer 200,000 square kilometers could be roofed over annually. While small in comparison to the approximately 150 million square kilometers of land area, consider that 200,000 1 kilometer sites scattered over the face of the Earth newly made habitable could revitalize vast swaths of land surrounding them—one square kilometer could grow local vegetables for a city in the desert, one over there could grow biofuel, enabling a desolate South Atlantic island to become fuel independent; at first, easily a billion people a year could be taken out of sweltering heat, biting cold and slashing rains, saving the money buying and running heating and air conditioning equipment would require.

Our design of the dome is presented in Fig. 3 that includes the thin inflated film dome. The innovations are listed here: (1) the construction is air-inflatable; (2) each dome is fabricated with very thin, transparent film (thickness is 0.1 to 0.3 mm) having controlled clarity and controlled heat conductivity without rigid supports; (3) the enclosing film has two conductivity layers plus a liquid crystal layer between them which changes its clarity, color and reflectivity under an electric voltage (fig. 4); (4) the bound section of dome has a hemispheric form (#5, fig.3). The air pressure is more in these sections and they protect the central sections from the outer wind.

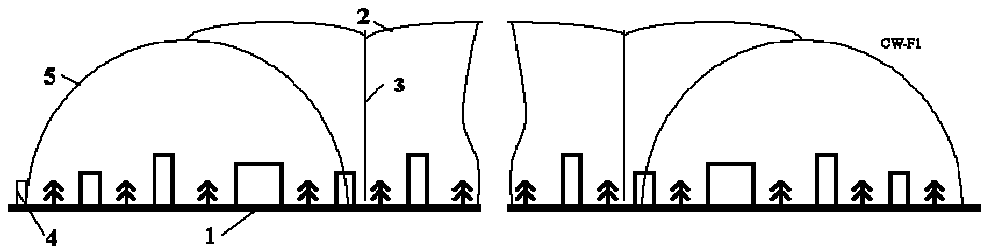

**Fig. 3.** Film dome over agriculture area or city. *Notations*: 1 - area, 2 - thin film cover with a control heat conductivity and clarity, 3 – control support cable and tubes for rain water ( height is 50 – 300 m), 4 - exits and ventilators, 5 - semi-cylindrical border section.

Fig. 3 illustrates the thin transparent control dome cover we envision. The inflated textile shell—technical "textiles" can be woven or non-woven (films)—embodies the innovations listed: (1) the film is very thin, approximately 0.1 to 0.3 mm., implying under 500 tons a square kilometer. A film this thin has never before been used in a major building; (2) the film has two strong nets, with a mesh of about $0.1 \times 0.1$ m and $a = 1 \times 1$ m, the threads are about 0.5 mm for a small mesh and about 1 mm for a big mesh. The net prevents the watertight and airtight film covering from being damaged by vibration; (3) the film incorporates a tiny electrically conductive wire net with a mesh about 0.1 x 0.1 m and a line width of about 100 $\mu$ and a thickness near 10 $\mu$. The wire net is electric (voltage) control conductor. It can inform the dome maintenance engineers concerning the place and size of film damage (tears, rips, etc.); (4) the film may be twin-layered with the gap — $c = 1$ m and $b = 2$ m—between film layers for heat insulation. In polar (and hot) regions this multi-layered covering is the main means for heat isolation and puncture of one of the layers won't cause a loss of shape because the second film layer is unaffected by holing; (5) the airspace in the dome's covering can be partitioned, either hermetically or not; and (6) part of the covering can have a very thin shiny aluminum coating that is about $1\mu$ (micron) for reflection of unnecessary solar radiation in equatorial or collect additional solar radiation in the polar regions [1].

The author offers a method for moving off the snow and ice from the film in polar regions. After snowfall we decrease the heat cover protection, heating the show (ice) by warm air flowing into channels 5 (fig.4) (between cover layers), and water runs down in tubes 3 (fig.3).

The town cover may be used as a screen for projecting of pictures, films and advertising on the cover at night times.

**Brif information about cover film.** Our dome cover (film) has 5 layers (fig. 4c): transparent dielectric layer, conducting layer (about 1 - 3 $\mu$), liquid crystal layer (about 10 - 100 $\mu$), conducting layer (for example, $SnO_2$), and transparant dielectric layer. Common thickness is 0.1 - 0.5 mm. Control voltage is 5 - 10 V. This film may be produced by industry relatively cheaply.

**1. Liquid crystals** (LC) are substances that exhibit a phase of matter that has properties between those of a conventional liquid, and those of a solid crystal.

Liquid crystals find wide use in liquid crystal displays (LCD), which rely on the optical properties of certain liquid crystalline molecules in the presence or absence of an electric field. The electric field can be used to make a pixel switch between clear or dark on command. Color LCD systems use the same technique, with color filters used to generate red, green, and blue pixels. Similar principles can be used to make other liquid crystal based optical devices. Liquid crystal in fluid form is used to detect electrically generated hot spots for failure analysis in the semiconductor industry. Liquid crystal memory units with extensive capacity were used in Space Shuttle navigation equipment. It is also worth noting that many common fluids are in fact liquid crystals. Soap, for instance, is a liquid crystal, and forms a variety of LC phases depending on its concentration in water.

The conventional control clarity (transparancy) film reflected a superfluos energy back to space. If film has solar cells that converts the superfluos solar energy into electricity.

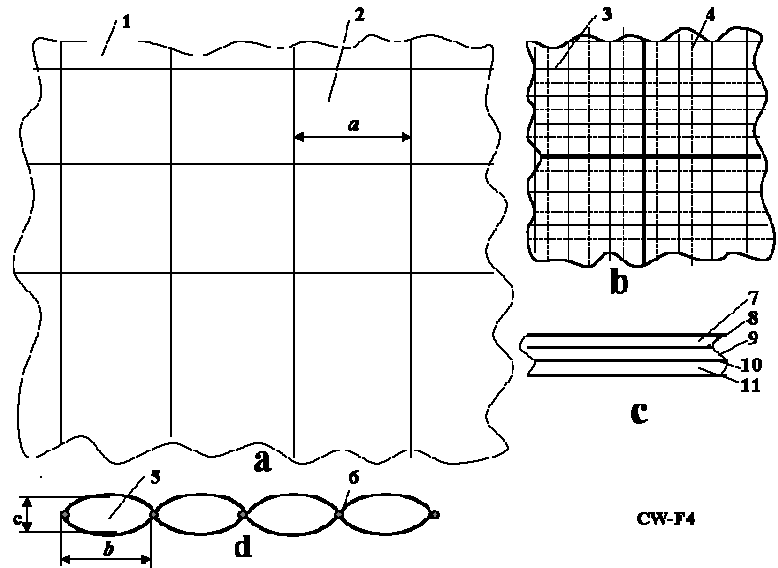

**Fig. 4.** Design of covering membrane. *Notations*: (a) Big fragment of cover with control clarity (reflectivity, carrying capacity) and heat conductivity; (b) Small fragment of cover; (c) Cross-section of cover (film) having 5 layers; (d) Longitudinal cross-section of cover for cold and hot regions; 1 - cover; 2 - mesh; 3 - small mesh; 4 - thin electric net; 5 - cell of cover; 6 - tubes;: 7 - transparant dielectric layer, 8 - conducting layer (about 1 - 3 μ), 9 - liquid crystal layer (about 10 - 100 μ), 10 - conducting layer, and 11 - transparant dielectric layer. Common thickness is 0.1 - 0.5 mm. Control voltage is 5 - 10 V.

**2. Transparency**. In optics, transparency is the material property of allowing light to pass through. Though transparency usually refers to visible light in common usage, it may correctly be used to refer to any type of radiation. Examples of transparent materials are air and some other gases, liquids such as water, most glasses, and plastics such as Perspex and Pyrex. Where the degree of transparency varies according to the wavelength of the light. From electrodynamics it results that only a vacuum is really transparent in the strict meaning, any matter has a certain absorption for electromagnetic waves. There are transparent glass walls that can be made opaque by the application of an electric charge, a technology known as electrochromics. Certain crystals are transparent because there are straight lines through the crystal structure. Light passes unobstructed along these lines. There is a complicated theory "predicting" (calculating) absorption and its spectral dependence of different materials.

**3. Electrochromism** is the phenomenon displayed by some chemical species of reversibly changing color when a burst of charge is applied.

One good example of an electrochromic material is polyaniline which can be formed either by the electrochemical or chemical oxidation of aniline. If an electrode is immersed in hydrochloric acid which contains a small concentration of aniline, then a film of polyaniline can be grown on the electrode. Depending on the redox state, polyaniline can either be pale yellow or dark green/black. Other electrochromic materials that have found technological application include the viologens and polyoxotungstates. Other electrochromic materials include tungsten oxide ($WO_3$), which is the main chemical used in the production of electrochromic windows or smart windows.

As the color change is persistent and energy need only be applied to effect a change,

electrochromic materials are used to control the amount of light and heat allowed to pass through windows ("smart windows"), and has also been applied in the automobile industry to automatically tint rear-view mirrors in various lighting conditions. Viologen is used in conjunction with titanium dioxide ($TiO_2$) in the creation of small digital displays. It is hoped that these will replace LCDs as the viologen (which is typically dark blue) has a high contrast to the bright color of the titanium white, therefore providing a high visibility of the display.

### 3. THEORY AND COMPUTATIONS of THE AB DOME

**1. General theory.**

As wind flows over and around a fully exposed, nearly completely sealed inflated dome, the weather affecting the external film on the windward side must endure positive air pressures as the wind stagnates. Simultaneously, low air pressure eddies will be present on the leeward side of the dome. In other words, air pressure gradients caused by air density differences on different parts of the dome's envelope is characterized as the "buoyancy effect". The buoyancy effect will be greatest during the coldest weather when the dome is heated and the temperature difference between its interior and exterior are greatest. In extremely cold climates such as the Arctic and Antarctic Regions the buoyancy effect tends to dominate dome pressurization.

Our basic computed equations, below, are derived from a Russian-language textbook [13]. Solar radiation impinging the orbiting Earth is approximately 1400 W/m². The average Earth reflection by clouds and the sub-aerial surfaces (water, ice and land) is about 0.3. The Earth-atmosphere absorbs about 0.2 of the Sun's radiation. That means about $q_0$ = 700 W/m²s of solar energy (heat) reaches our planet's surface in cloudy weather at the Equator. That means we can absorb about 30 - 80% of solar energy. It is enough for normal plant growth in wintertime (up to 40-50° latitude) and in circumpolar regions with a special variant of the dome design.

The solar spectrum is graphically portrayed in Fig. 5.

The visible part of the Sun's spectrum is only $\lambda$ = 400 – 800 nm (0.4 to 0.8 $\mu$.). Any warm body emits radiation. The emission wavelength depends on the body's temperature. The wavelength of the maximum intensity (see Fig. 5) is governed by the black-body law originated by Max Planck (1858-1947):

$$\lambda_m = \frac{2.9}{T}, \quad [mm], \qquad (1)$$

where $T$ is body temperature, °K. For example, if a body has an ideal temperature 20 °C ($T$ = 293 °K), the wavelength is $\lambda_m$ = 9.9 $\mu$.

The energy emitted by a body may be computed by employment of the Josef Stefan-Ludwig Boltzmann law:

$$E = \varepsilon \sigma_s T^4, \quad [W/m^2], \qquad (2)$$

where $\varepsilon$ is coefficient of body blackness ($\varepsilon$ =0.03 ÷ 0.99 for real bodies), $\sigma_s$ = 5.67×10$^{-8}$ [W/m²·K] Stefan-Boltzmann constant. For example, the absolute black-body ($\varepsilon$ = 1) emits (at $T$ = 293 °K) the energy $E$ = 418 W/m².

**2. Cold regions.**

Amount of the maximum solar heat flow at 1 m² per 1 second of Earth surface is

$$q = q_o \cos(\varphi \pm \theta) \quad [W/m^2], \qquad (3)$$

where $\varphi$ < 90° is Earth longevity, $\theta$ < 23.5° is angle between projection of Earth polar axis to the plate which is perpendicular to the ecliptic plate and contains the line Sun-Earth and the perpendicular to ecliptic plate. The sign "+" signifies Summer and the "-" signifies Winter, $q_o \approx$ 700 W/m² is the annual average solar heat flow to Earth at equator corrected for Earth reflectance. For our case this magnitude can reach $q_o \approx$ 1000 - 1100 W/m² or in clouded sky the magnitude decreases up 100 W/m².

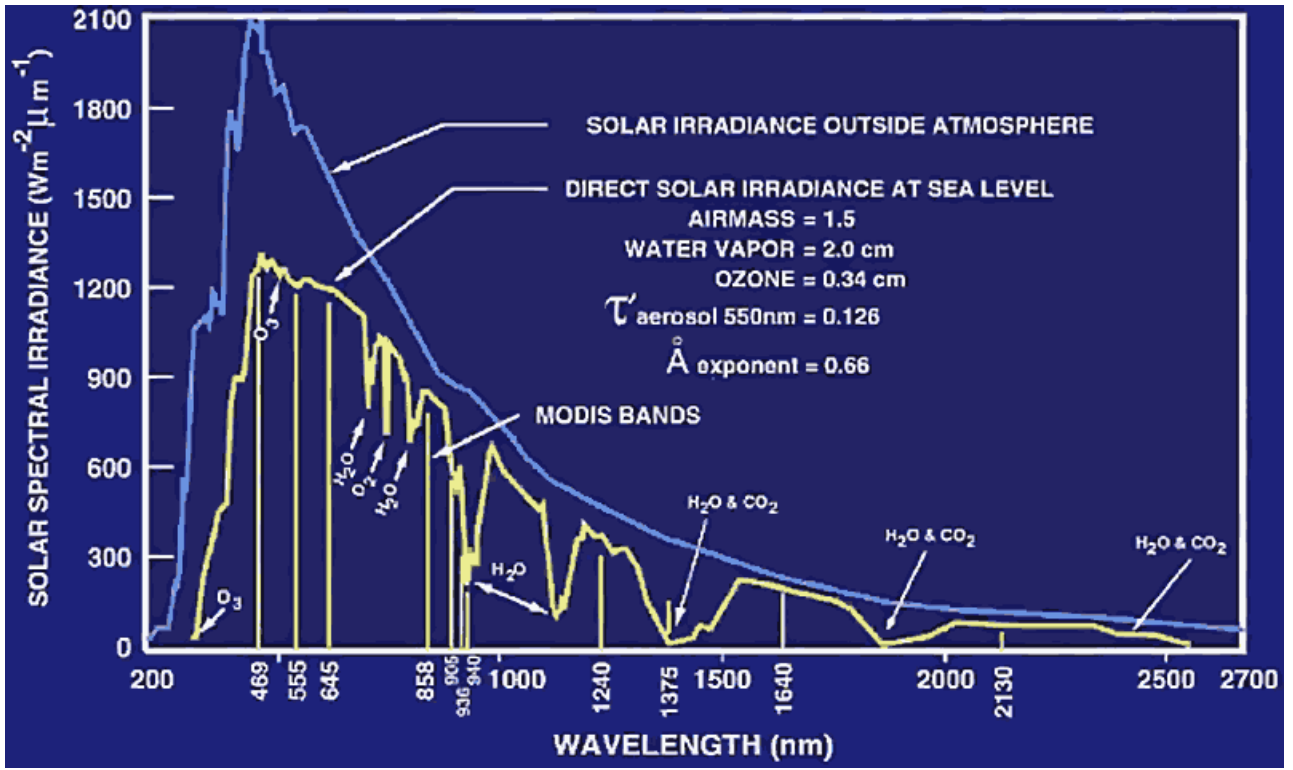

**Fig.5.** Spectrum of solar radiation. Visible light is 400 - 800 nm.

This angle is changed during a year and may be estimated for Earth's North Polar Region hemisphere by the following the first approximation equation:

$$\theta = \theta_m \cos\omega, \quad \text{where} \quad \omega = 2\pi \frac{N}{364}, \tag{4}$$

where $\theta_m$ is maximum $\theta$, $|\theta_m| = 23.5° = 0.41$ radian; $N$ is number of day in a year. The computations for summer and winter are presented in fig.6.

The heat flow for a hemisphere having reflector [1] at noon may be computed by the equation:

$$q = c_1 q_0 [\cos(\varphi - \theta) + S \sin(\varphi + \theta)], \tag{5}$$

where $S$ is fraction (relative) area of reflector to service area of "Evergreen" dome [1]. For reflector of Fig.1 [1] $S = 0.5$; $c_1$ is film transparency coefficient ($c_1 \approx 0.8 - 0.95$).

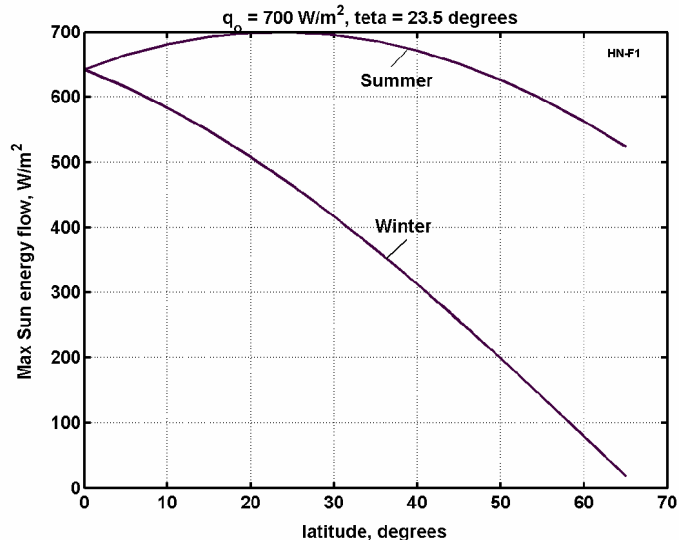

**Fig.6.** Maximum Sun radiation flow at Earth surface via Earth latitude and season without dome.

The daily average solar irradiation (energy) is calculated by equation

$$Q = 86400\, c\, q\, t, \quad \text{where} \quad t = 0.5(1 + \tan\varphi \tan\theta), \quad |\tan\varphi \tan\theta| \leq 1, \qquad (6)$$

where $c$ is daily average heat flow coefficient, $c \approx 0.5$ without dome, $c \approx 0.75$ with dome; $t$ is relative daily illuminated time, $86400 = 24 \times 60 \times 60$ is the number of seconds in a day.
The computation for relative daily light time is presented in Fig. 7.

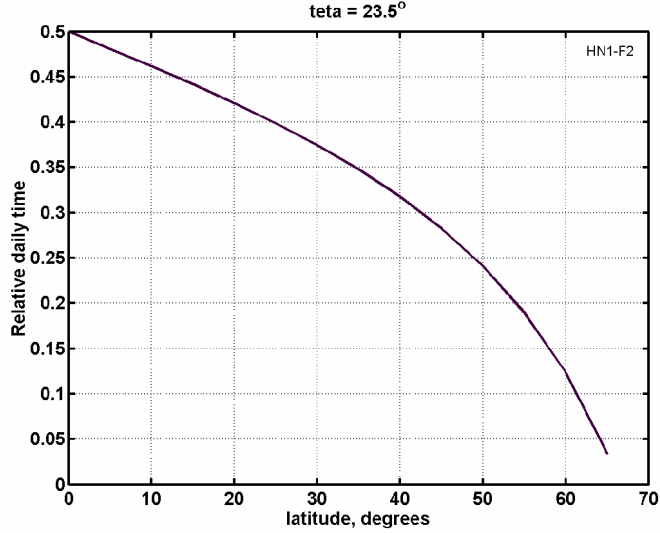

**Fig.7**. Relative daily light time via Earth latitude.

The convective (conductivity) heat flow per 1 m$^2$ of dome film cover by convection and heat conduction is (see [2]):

$$q = k(t_1 - t_2), \quad \text{where} \quad k = \frac{1}{1/\alpha_1 + \sum_i \delta_i / \lambda_i + 1/\alpha_2}, \qquad (7)$$

where $k$ is heat transfer coefficient, W/m$^2$·K; $t_{1,2}$ are temperatures of the inter and outer multi-layers of the heat insulators, °C; $\alpha_{1,2}$ are convention coefficients of the inter and outer multi-layers of heat insulators ($\alpha = 30 \div 100$), W/m$^2$K; $\delta_i$ are thickness of insulator layers; $\lambda_i$ are coefficients of heat transfer of insulator layers (see Table 1), m; $t_{1,2}$ are temperatures of initial and final layers °C.

The radiation heat flow per 1 m$^2$ of the service area computed by equations (2):

$$q = C_r\left[\left(\frac{T_1}{100}\right)^4 - \left(\frac{T_2}{100}\right)^4\right], \quad \text{where} \quad C_r = \frac{c_s}{1/\varepsilon_1 + 1/\varepsilon_2 - 1}, \quad c_s = 5.67 \; [\text{W/m}^2\text{K}^4], \qquad (8)$$

where $C_r$ is general radiation coefficient, $\varepsilon$ are black body rate (emittance) of plates (see Table 2); $T$ is temperatures of plates (surface), °K.

The radiation flow across a set of the heat reflector plates is computed by the equation:

$$q = 0.5 \frac{C'_r}{C_r} q_r, \qquad (9)$$

where $C'_r$ is computed by equation (8) between plate and reflector.
The data of some construction materials is found in Table 3, 4.

**Table 3**. [13], p.331. **Heat Transfer Data**

| Material | Density, kg/m$^3$ | Thermal conductivity, $\lambda$, W/m·°C | Heat capacity, kJ/kg·°C |
|----------|-------------------|------------------------------------------|--------------------------|
| Concrete | 2300              | 1.279                                    | 1.13                     |
| Baked brick | 1800 | 0.758 | 0.879 |
| Ice | 920 | 2.25 | 2.26 |
| Snow | 560 | 0.465 | 2.09 |
| Glass | 2500 | 0.744 | 0.67 |
| Steel | 7900 | 45 | 0.461 |
| Air | 1.225 | 0.0244 | 1 |

---

As the reader will see, the air layer is the best heat insulator. We do not limit its thickness $\delta$.

**Table 4**. [13], p. 465. Emittance, ε **(Emissivity)**

| Material | Temperature, $T\,°C$ | Emittance, ε |
|---|---|---|
| Bright Aluminum | $50 \div 500\,°C$ | 0.04 - 0.06 |
| Bright copper | $20 \div 350\,°C$ | 0.02 |
| Steel | $50\,°C$ | 0.56 |
| Asbestos board | $20\,°C$ | 0.96 |
| Glass | $20 \div 100\,°C$ | 0.91 - 0.94 |
| Baked brick | $20\,°C$ | 0.88 - 0.93 |
| Tree | $20\,°C$ | 0.8 - 0.9 |
| Black vanish | $40 \div 100\,°C$ | 0.96 – 0.98 |
| Tin | $20\,°C$ | 0.28 |

As the reader will notice, the shiny aluminum louver coating is an excellent means of retention against radiation losses from the dome.

The general radiation heat $Q$ computes by equation (6). Equations (1) – (9) allow computation of the heat balance and comparison of incoming heat (gain) and outgoing heat (loss).

The computations of heat balance of a dome (with reflector mirror [1]) of any size in the coldest wintertime of the Polar Regions are presented in Fig. 8.

The thickness of the dome envelope, its sheltering shell of film, is computed by formulas (from equation for tensile strength):

$$\delta_1 = \frac{Rp}{2\sigma}, \quad \delta_2 = \frac{Rp}{\sigma}, \qquad (10)$$

where $\delta_1$ is the film thickness for a spherical dome, m; $\delta_2$ is the film thickness for a cylindrical dome, m; $R$ is radius of dome, m; $p$ is additional pressure into the dome (10÷1000), N/m$^2$; $\sigma$ is safety tensile stress of film (up to $2\times10^9$), N/m$^2$.

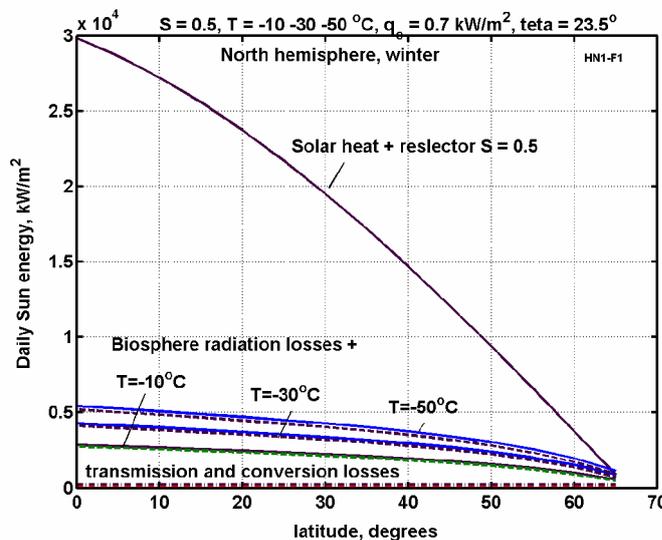

**Fig. 8.** Daily heat balance through 1 m$^2$ of dome with mirror during coldest winter day versus Earth's latitude (North hemisphere example). Data used for computations (see Eq. (1) - (9)): temperature inside

of dome is $t_1 = +20\,°C$, outside are $t_2 = -10, -30, -50\,°C$; reflectivity coefficient of mirror is $c_2 = 0.9$; coefficient transparency of film is $c_1 = 0.9$; convectively coefficients are $\alpha_1 = \alpha_2 = 30$; thickness of film layers are $\delta_1 = \delta_2 = 0.0001$ m; thickness of air layer is $\delta = 1$ m; coefficient of film heat transfer is $\lambda_1 = \lambda_3 = 0.75$, for air $\lambda_2 = 0.0244$; ratio of cover blackness $\varepsilon_1 = \varepsilon_3 = 0.9$, for louvers $\varepsilon_2 = 0.05$.

The dynamic pressure from wind is

$$p_w = \frac{\rho V^2}{2}, \tag{11}$$

where $\rho = 1.225$ kg/m$^3$ is air density; $V$ is wind speed, m/s.

For example, a storm wind with speed $V = 20$ m/s, standard air density is $\rho = 1.225$ kg/m$^3$. Then dynamic pressure is $p_w = 245$ N/m$^2$. That is four time less when internal pressure $p = 1000$ N/m$^2$. When the need arises, sometimes the internal pressure can be voluntarily decreased, bled off.

In Fig. 8 the alert reader has noticed: *the daily heat loss is about the solar heat in the very coldest winter day when a dome located above $60^0$ North or South Latitude and the outside air temperature is $-50\,°C$.*

In [1] we show the heat loss of the dome in Polar region is less than 14 times the heat of the buildings inside unprotected by an inflated dome.

We consider a two-layer dome film and one heat screen. If needed, better protection can further reduce the heat losses as we can utilize inflated dome covers with more layers and more heat screens. One heat screen decreases heat losses by 2, two screens can decrease heat flow by 3 times, three by 4 times, and so on. If the Polar Region domes have a mesh structure, the heat transfer decreases proportional to the summary thickness of its' enveloping film layers.

The dome shelter innovations outlined here can be practically applied to many climatic regimes (from Polar to Tropical). The North and South Poles may, during the 21$^{st}$ Century, become places of cargo and passenger congregation since the a Cable Space Transportation System, installed on Antarctica's ice-cap and on a floating artificial ice island has been proposed that would transfer people and cargoes to and from the Moon.[3]

### 3. Irrigation without water. Closed-loop water cycle.

A reader can derive the equations below from well-known physical laws [12]. Therefore, the author does not give detailed explanations of these.

1. **Amount of water in atmosphere**. Amount of water in atmosphere depends upon temperature and humidity. For relative humidity 100%, the maximum partial pressure of water vapor is shown in Table 5.
2. 

**Table 5**. Maximum partial pressure of water vapor in atmosphere for given air temperature

| $t$, C | -10 | 0 | 10 | 20 | 30 | 40 | 50 | 60 | 70 | 80 | 90 | 100 |
|---|---|---|---|---|---|---|---|---|---|---|---|---|
| $p$,kPa | 0.287 | 0.611 | 1.22 | 2.33 | 4.27 | 7.33 | 12.3 | 19.9 | 30.9 | 49.7 | 70.1 | 101 |

The amount of water in 1 m$^3$ of air may be computed by equation

$$m_W = 0.00625\,[p(t_2)h - p(t_1)], \tag{12}$$

where $m_W$ is mass of water, kg in 1 m$^3$ of air; $p(t)$ is vapor (steam) pressure from Table 5, relative $h = 0 \div 1$ is relative humidity. The computation of equation (12) is presented in fig.9. Typical relative humidity of atmosphere air is 0.5 - 1.

**Computation of closed-loop water cycle.** Assume the maximum safe temperature is achieved in the daytime. When dome reaches the maximum (or given) temperature, the control system fills with air the space 5 (fig.4) between double–layers of the film cover. That protects the inside part of the dome from further heating by outer (atmospheric) hot air. The control system decreases also the solar radiation input, increasing reflectivity of the liquid crystal layer of the film cover. That way, we can support a constant temperature inside the dome.

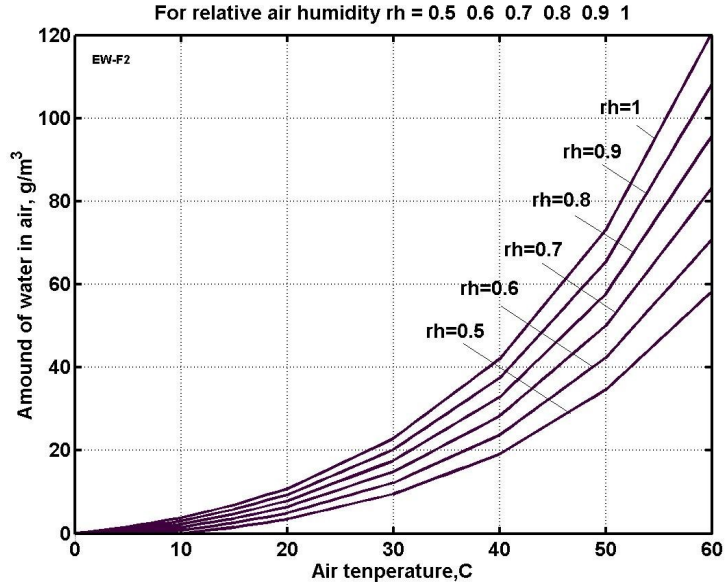

**Fig. 9**. Amount of water in 1 m$^3$ of air versus air temperature and relative humidity (rh). $t_1 = 0$ °C.

The **heating** of the dome in the daytime may be computed by equations:

$$q(t) = q_0 \sin(\pi t / t_d), \quad dQ = q(t)dt, \quad Q = \int_0^{t_d} dQ, \quad Q(0) = 0, \quad M_w = \int_0^{t_d} a dT,$$

$$dT = \frac{dQ}{C_{p1}\rho_1\delta_1 + C_{p2}\rho_2 H + rHa}, \quad a = 10^{-5}(5.28T + 2), \quad T = \int_0^{t_d} dT, \quad T(0) = T_{min}, \quad (13)$$

where $q$ is heat flow, J/m$^2$ s; $q_o$ is maximal Sun heat flow in noon daily time, $q_o \approx 100 \div 1000$, J/m$^2$s; $t$ is time, s; $t_d$ is daily (Sun) time, s; $Q$ is heat, J; $T$ is temperature in dome (air, soil), °C; $C_{p1}$ is heat capacity of soil, $C_{p1} \approx 1000$ J/kg; $C_{p2} \approx 1000$ J/kg is heat capacity of air; $\delta_1 \approx 0.1$ m is thickness of heating soil; $\rho_1 \approx 1000$ kg/m$^3$ is density of the soil; $\rho_2 \approx 1.225$ kg/m$^3$ is density of the air; $H$ is thickness of air (height of cover), $H \approx 50 \div 300$ m; $r = 2,260,000$ J/kg is evaporation heat, $a$ is coefficient of evaporation; $M_w$ is mass of evaporation water, kg/m$^3$; $T_{min}$ is minimal temperature into dome after night, °C.

The convective (conductive) **cooling** of dome at night time may be computed as below

$$q_t = k(T_{min} - T(t)), \quad \text{where} \quad k = \frac{1}{1/\alpha_1 + \sum_i \delta_i / \lambda_i + 1/\alpha_2} \quad (14)$$

where $q_t$ is heat flow through the dome cover by convective heat transfer, J/m$^2$s or W/m$^2$; see the other notation in Eq. (7). We take $\delta = 0$ in night time (through active control of the film).

The radiation heat flow (from dome to night sky, radiation cooling) may be estimated by equations:

$$q_r = C_r \left[\left(\frac{T_{min}}{100}\right)^4 - \left(\frac{T(t)}{100}\right)^4\right], \quad \text{where} \quad C_r = \frac{c_s}{1/\varepsilon_1 + 1/\varepsilon_2 - 1}, \quad c_s = 5.67 \text{ [W/m}^2\text{K}^4\text{]}, \quad (15)$$

where $q_r$ is heat flow through dome cover by radiation heat transfer, J/m$^2$s or W/m$^2$; see the other notation in Eq. (8). We take $\varepsilon = 1$ in night time (through active control of the film).

The other equations are same (13)

$$dQ = [q_t(t) + q_r(t)]dt, \quad Q = \int_0^{t_d} dQ, \quad Q(0) = 0, \quad M_w = \int_0^{t_d} a\,dT,$$

$$dT = \frac{dQ}{C_{p1}\rho_1\delta_1 + C_{p2}\rho_2 H + rHa}, \quad a = 10^{-5}(5{,}28T + 2), \quad T = \int_0^{t_d} dT, \quad T(0) = T_{min}, \quad (16)$$

Let us take the following parameters: $H = 135$ m, $\alpha = 70$, $\delta = 1$ m between cover layers (see #5 in Fig.4), $\lambda = 0.0244$ for air. Result of computation for given parameter are presented in figs. 10 – 13.

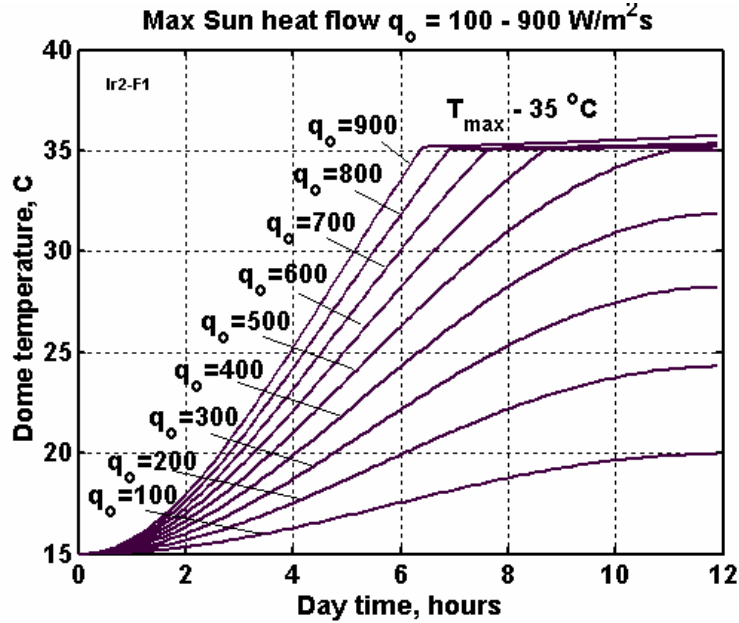

**Fig. 10.** Heating of the dome by solar radiation from the night temperature of $15\,^\circ$C to $35\,^\circ$C via daily maximal solar radiation (W/m$^2$) for varying daily time. Height of dome film cover equals $H = 135$ m. The control temperature system limits the maximum internal dome temperature to $35\,^\circ$C. Compare with Fig.6. *You see the offered AB Dome can support the average daily temperature of $18\,^\circ$C in winter time up to latitude $50^\circ$.*

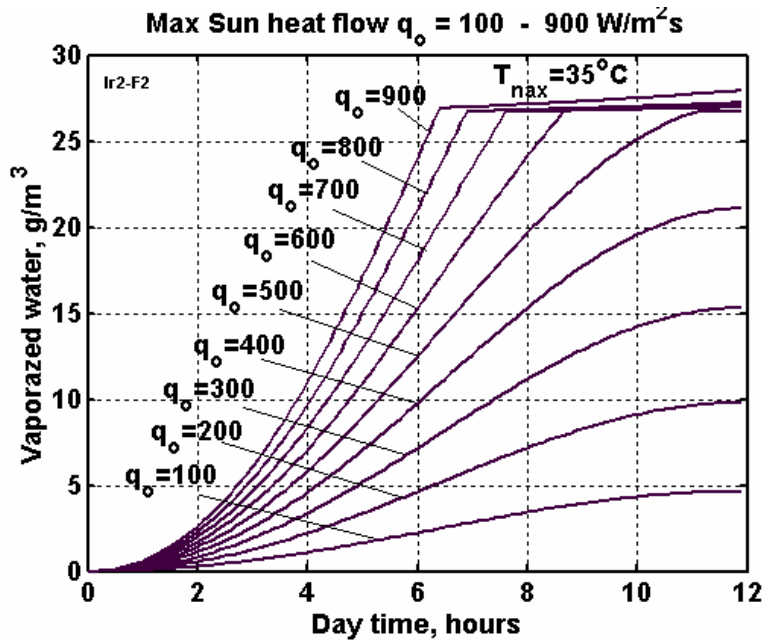

**Fig.11.** Water vaporization for 100% humidity of the air for different maximal solar radiation (W/m$^2$) levels delivered over varying daily time. Height of dome film cover equals $H = 135$ m. The temperature control system limits the maximum internal dome temperature to 35 C.

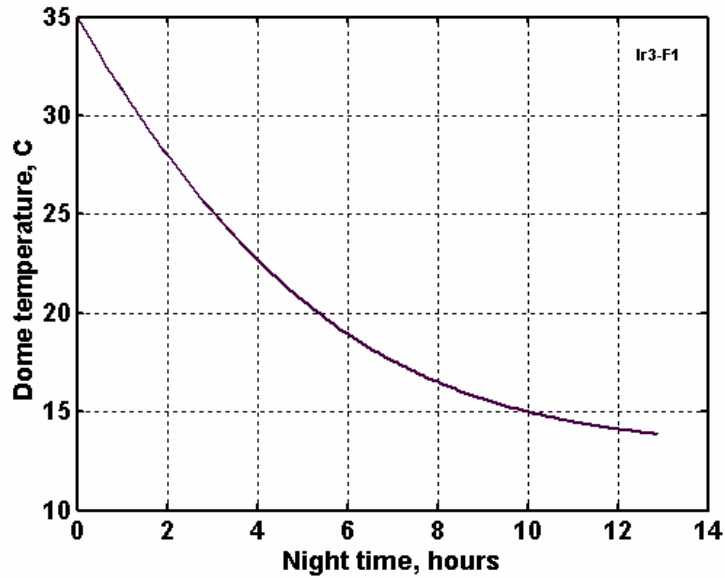

**Fig.12.** Cooling of the Dome via night time for initial daily temperature 35 º C and the night outer temperature 13 º C.

For dome cover height $H$ = 135 m the night precipitation (maximum) is 0.027×135 = 3.67 kg (liter) or 3.67 mm/day (Fig.11). The annual precipitation is 1336.6 mm (maximum). If it is not enough, we can increase the height of dome cover. The globally-averaged annual precipitation is about 1000 mm.

As you see, we can support the same needed temperature in a wide range of latitudes at summer and winter time. That means the covered regions are not hostage to their location upon the Earth's surface (up to latitude 40 º -50º), nor Earth's seasons, or the **vagaries** of outside weather. Our design of Dome is not optimal, but rather selected for realistic parameters.

### 4. DISCUSSION

As with any innovative macro-project proposal, the reader will naturally have many questions. We offer brief answers to the four most obvious questions our readers are likely to ponder.
(1) *How can snow and ice be removed from the dome?*
If water appears over film (rain), it flows down through a special tube. If snow (ice) appears atop the film, the control system passes the warm air between two cover layers. The warm air melts the snow (ice) and water flows down. The film cover is flexible and has a lift force of about 10 -100 kg/m².
(2) *Storm wind.*
The storm wind can only be on the bounding (outside) sections of dome. They are special semi-cylindrical form (fig.3) and stronger than central sections.
(3) *Cover damage.*
The envelope contains a rip-stop cable mesh so that the film cannot be damaged greatly. Electronic signals alert supervising personnel of any rupture problems. The needed part of cover may be reeled down by control cable and repaired.
(3) *What is the design life of the film covering?*
Depending on the kind of materials used, it may be as much a decade. In all or in part, the cover can be replaced periodically.

# 5. CONCLUSION

One half of Earth's population is malnourished. *The majority of Earth is not suitable for unshielded human life*. The increasing of agriculture area, crop capacity, carrying capacity by means of converting the deserts, desolate wilderness, taiga, permafrost into gardens are an important escape hatch from some of humanity's most pressing problems. The offered cheapest ($0.1 ÷ 0.3/m$^2$) AB method may dramatically increase the potentially realizable sown area, crop capacity; indeed the range of territory suitable for human living. In theory, converting all Earth land such as Alaska, North Canada, Siberia, or the Sahara or Gobi deserts into prosperous garden would be the equivalent of colonizing an entire new planet. The suggested method is very cheap (cost of covering 1 m$^2$ is about 10 - 30 cents) and may be utilized at the present time. We can start from small areas, such as small towns in bad regions and extended the practice over a large area—and what is as important, making money most of the way.

Film domes can foster the fuller economic development of dry, hot, and cold regions such as the Earth's Arctic and Antarctic and, thus, increase the effective area of territory dominated by humans. Normal human health can be maintained by ingestion of locally grown fresh vegetables and healthful "outdoor" exercise. The domes can also be used in the Tropics and Temperate Zone. Eventually, they may find application on the Moon or Mars since a vertical variant, inflatable towers to outer space, are soon to become available for launching spacecraft inexpensively into Earth-orbit or interplanetary flights [12].

The related problems are researched in references [1]-[12].

Let us shortly summarize some advantages of this offered AB Dome method of climate moderation:

(1) Method does not need large amounts of constant input water for irrigation;
(2) Low cost of inflatable film Dome per area reclaimed: (10 - 30 cents/m$^2$);
(3) Control of inside temperature;
(4) Usable in very hot and cool regions;
(5) Covered area is not at risk from weather;
(6) Possibility of flourishing crops even with a sterile soil (hydroponics);
(7) 2 – 3 harvests in year; without farmers' extreme normal risks.
(8) Rich harvests, at that.
(9) Converting deserts, desolate wilderness, taiga, tundra, permafrost, and ocean into gardens;
(10) Covering the towns, cities by offered domes;
(11) Using the dome cover for illumination, pictures, films and advertising.

We can make fine local weather, get new territory for living with an agreeable climate without daily rain, wind and low temperatures, and for agriculture. We can cover by thin film gigantic expanses of bad dry and cold regions. The countries having big territory (but bad land) may be able to use to increase their population and became powerful states in the centuries to come.

The offered method may be used to conserve a vanishing sea as the Aral or Dead Sea. A closed loop water cycle saves this sea for a future generation, instead of bequeathing a salty dustbowl.

The author developed the same method for the ocean (sea). By controlling the dynamics and climate there, ocean colonies may increase the useful area another 3 times (after the doubling of useful land outlined above) All in all, this method would allow increasing the Earth's population by 5 – 10 times without the starvation.

The offered method can solve the problem of **global warming** because AB domes will be able to confine until use much carbonic acid ($CO_2$) gas which appreciably increases a harvest. This carbon will show up in yet more productive crops! The dome lift force reaches up 300 kg/m$^2$. The telephone, TV, electric, water and other communications can be suspended to the dome cover.

The offered method can also help to defend the cities (or an entire given region) from rockets, nuclear warheads, and military aviation. Details may be offered in a later paper.

## Acknowledgement

The author wishes to acknowledge Joseph Friedlander for correcting the author's English.